\documentstyle[mprocl]{article}

\def\Journal#1#2#3#4{{#1} {\bf #2}, #3 (#4)}

\def\NCA{{\em Nuovo Cimento}}

\def\PRD{{\em Phys. Rev.} D}

\def\be{\begin{equation}}
\def\ee{\end{equation}}
\def\bea{\begin{eqnarray}}
\def\eea{\end{eqnarray}}

\newcommand{\del}{\delta}

\def\footnote{\def\dummy}  

\begin{document}

\title{
       On the perturbation of domain wall coupled to 
       gravitational waves   
      }

\author{Akihiro ISHIBASHI and Hideki ISHIHARA }

\address{
        Department of Physics, Tokyo Institute of Technology, \\ 
        Oh-Okayama Meguro-ku, Tokyo 152, Japan
         }


\maketitle

\abstracts{
         The equation of motion for domain wall coupled to gravitational 
         field is derived. The domain wall is treated as a source 
         of gravitational field around the wall. 
         The perturbed equation is also obtained with taking account of 
         the gravitational back reaction on the motion of the domain wall. 
         }
  


  As a possible evidence of cosmological phase transitions in the early 
universe, topological defects may remain somewhere in our universe~\cite{VS}. 
It has been of interest to investigate their behaviors, especially, 
those accompanied with radiations in cosmological and astrophysical 
contexts. We shall comment on the equation of motion (EOM) for domain wall 
coupled to gravity, especially the perturbed EOM for domain wall 
coupled to gravitational waves.  


In the treatment of domain walls, we often apply the thin wall approximation. 
Then the history of the domain wall is regarded as a 3-dimensional timelike 
hypersurface $(\Sigma, \gamma_{ij})$ in 4-dimensional spacetime 
$(M,g_{\mu \nu})$ embedded by $x^\mu = x^\mu (\zeta^j)$, 
where the 3-dimensional metric is induced by 
$ \gamma_{ij} := g_{\mu \nu} ({\partial x^{\mu}} / {\partial \zeta^{i}})
                           ({\partial x^{\nu}} / {\partial \zeta^{j}})
$. 
The EOM for the domain wall is derived from the Nambu-Goto action 
\be 
S = - \sigma \int_{\Sigma} d^3\zeta \sqrt{- \gamma} 
    + \rho \int_{M_{inside}}d^4 x \sqrt{-g}, 
\label{action}
\ee 
where $\sigma$ is the surface energy density of the wall and 
$\rho$ is the difference of the vacuum energy densities 
between the inside and the outside regions of the wall. 
Taking the variation in the wall position $x^\mu$ gives the EOM, 
$
   K = - {\rho} / {\sigma}
$, 
where $K$ is the trace of the extrinsic curvature $K_{ij}$ of $\Sigma$. 


The perturbation of the domain wall is described by 
$ 
\tilde{x}^\mu = x^\mu + \phi n^\mu
$, 
where $n^\mu$ is the normal vector to $\Sigma$. 
The scalar field $\phi$ living on $\Sigma$ obeys the Klein-Gordon equation, 
\be 
   \left( \Box_3 - m^2 \right) \phi = 0, 
\label{KG}
\ee 
on the background world sheet of the wall~\cite{GV,G}. 
Here $ \Box_3$ is the d'Alembertian for $\gamma_{ij}$ and 
\be 
    {m^2} := - {R}_{\mu \nu}q^{\mu \nu} + {}^{(3)}R 
                     - \left( \frac{\rho}{\sigma} \right)^2, 
\ee 
where $ R_{\mu \nu} $ is the spacetime Ricci tensor evaluated 
at the world hypersheet and ${}^{(3)}R$ the 3-dimensional scalar curvature of 
$(\Sigma, \gamma_{ij})$. 
From Eq.~(\ref{KG}), one intuitively expect that 
the wave propagation of the scalar field represents the oscillation 
of the deformed wall and such wall oscillations generate 
gravitational waves. 

In the derivation of Eq.~(\ref{KG}) 
one naively assumed that the gravitational back reaction 
on the wall motion was higher order 
and hence the influence of gravitational waves could be negligible 
at least in the first order of wall perturbation. 


However, recent investigations of the perturbations of O(3,1) symmetric
domain walls~\cite{KIF,II} demonstrate 
that the naive assumption noted above is incorrect. 
In these works~\cite{KIF,II}, by the metric junction formalism, 
the moving domain wall is treated as a source of the gravitational 
field. It turns out that the perturbative motion of the wall 
is completely accompanied with the gravitational perturbations 
and the domain wall oscillates only while incident gravitational waves 
go across; the domain wall loses its own dynamical degree of freedom. 
Thus, the gravitational back reaction on a domain wall motion 
cannot be ignored even in the first order of the perturbation. 


Motivated by these works, we shall consider how to derive 
the perturbed EOM for domain wall coupled to gravitational waves 
from the Nambu-Goto action with the action for the gravitational field  
$ \int d^4 x \sqrt{- g} R / {16 \pi G} $. 

Note that, once the wall $\Sigma$ is regarded as a source of 
a gravity, $\Sigma$ is a singular hypersurface and 
the spacetime metric $g_{\mu \nu}$ is no longer smooth there. 
Thus, 
while the intrinsic quantities are continuous and well-defined at $\Sigma$, 
the derivatives of the spacetime metric such as the extrinsic curvature 
$K_{ij}$ and the spacetime Ricci tensor $R_{\mu \nu}$ are 
in general discontinuous at $\Sigma$ and ill-defined on $\Sigma$. 
The EOM therefore has ambiguities in the evaluations of 
the discontinuous quantities at $\Sigma$. 

To resolve this problem, we evaluate the discontinuous quantities on $\Sigma$ 
to be consistent with the Einstein equations 
on the whole spacetime~$(M, g_{\mu \nu})$. 
Since the $(i,j)$ components of the Einstein tensor $G{}^i{}_j$ contains 
$
\partial_\chi K{}^i{}_j, 
$ 
where $\chi$ is the axis normal to $\Sigma$, 
and the energy momentum tensor $T{}^\mu{}_\nu$ contains the delta function, 
one obtain 
$
    K{}^i{}_j|_\Sigma 
    := 
        \lim_{\epsilon \rightarrow 0} 
            \int^{\epsilon}_{- \epsilon} d\chi 
                                              K{}^i{}_j \del (\chi) 
    = 
       \frac{1}{2}
               \lim_{\epsilon \to 0} 
                        \left\{ 
                                K{}^i{}_j|_{+ \epsilon} 
                                + K{}^i{}_j|_{- \epsilon} 
                        \right\} 
    =: 
       \overline{K}{}^i{}_j
$. 
Similarly, for $R_{ij}$, 
we should consider the averaged one $\overline{R_{ij}}$. 
This is indeed what the metric junction formalism yields. 
Then we obtain the EOM for a domain wall coupled to gravitational field 
\be 
  \overline{K} = - {\rho}/{\sigma}. 
\label{eq:WM}  
\ee       


Now let us consider the perturbed equation. 
To study a coupled system of a domain wall and gravitational field, 
in addition to the wall perturbation $\phi$, 
we further take the metric perturbations 
$\del g_{\mu \nu} =: h_{\mu \nu}$ into account. 
Then we see that the spacetime metric transforms as  
$  
g_{\mu \nu}(x) \longrightarrow 
     g_{\mu \nu}(x) - 2\phi K_{\mu \nu}(x) + h_{\mu \nu}(x)
$. 
Taking the average for the discontinuous quantities, 
we obtain the perturbed EOM for the domain wall coupled to gravitational 
perturbations in the form 
\be 
 \left(
        \Box_3  - \overline{m^2}
 \right) {\phi (\zeta^i)}  
         + {J} = 0, 
\label{Eom:3dim}
\ee 
where 
\be 
     J := {}^{(3)}\!D^j  h_{\chi j} 
              - \frac{1}{2} \overline{ \partial_{\chi} h^j{}_j } 
              + \frac{1}{2}
                           \left(
                                 {\rho}/{\sigma} 
                           \right) 
                                  h_{\chi \chi}. 
\ee 


In particular spherically symmetric background case, 
introducing the gauge-invariant variables for the metric perturbations 
following Gerlach and Sengupta~\cite{GS}, 
we can express Eq.~(\ref{Eom:3dim}) in terms of them as 
     \be 
        \left( \Box_3 - \overline{m^2} \right) \Xi + J= 0, 
     \ee 
where 
\be 
   \Xi := \phi + X 
\ee 
is the gauge-invariant wall displacement variable and 
$X$ is a quantity corresponding to $n^C P_C$ introduced by 
Gerlach and Sengupta~\cite{GS}. 
For the detail of the derivation, see the paper~\cite{II2}. 


\end{document}